\documentclass[12pt]{article}
\usepackage[psamsfonts]{euscript}
\usepackage[psamsfonts]{amssymb}
\usepackage{amsmath}
\usepackage{graphicx}
\usepackage[bf,small,center]{titlesec}
\usepackage[small]{caption}

\newcommand{\nc}{\newcommand}
\nc{\carb} {{}^{12}\mathrm{C}}
\nc{\dL}    {\Delta_\mathrm{L}}
\nc{\dmh}   {\dot M_\mathrm{H}}
\nc{\dmhe}  {\dot M_{3\alpha}}
\nc{\fh}   {f_\mathrm{H}}
\nc{\fhe}  {f_{3\alpha}}
\nc{\hel}  {{}^4\mathrm{He}}
\nc{\hydr} {{}^1\mathrm{H}}
\nc{\K}    {\:\mathrm{K}}
\nc{\Lbol} {L_\mathrm{bol}}
\nc{\mzams}{M_\mathrm{ZAMS}}
\nc{\ox}   {{}^{16}\mathrm{O}}
\nc{\Teff} {T_\mathrm{eff}}
\nc{\ts}   {t_\mathrm{s}}
\nc{\umax} {U_{\max}}
\nc{\vesc} {v_\mathrm{esc}}
\begin{document}

\begin{center}
\textbf{
MASS--LUMINOSITY RELATION AND \\[4pt]
PULSATIONAL PROPERTIES OF WOLF--RAYET STARS
}

\vskip 3mm
\textbf{Yu.A. Fadeyev\footnote{e--mail: fadeyev@inasan.ru}}

\textit{Institute of Astronomy of the Russian Academy of Sciences, Moscow} \\

Submitted 12 May 2008
\end{center}

\vskip 10pt\noindent
Evolution of Population~I stars with initial masses ranging within
$70M_\odot\le\mzams\le 130M_\odot$ is considered under various
assumptions on the mass loss rate $\dot M$.
The mass--luminosity relation of W--R stars is shown to be most
sensitive to the mass loss rate $\dmhe$ during the helium burning phase.
Together with the mass--luminosity relation obtained for all
evolutionary sequences several more exact relations
are determined for the constant ratios $0.5\le \fhe\le 3$,
where $\fhe = \dot M/\dmhe$.
Evolutionary models of W--R stars were used as initial conditions in
hydrodynamic computations of radial nonlinear stellar oscillations.
The oscillation amplitude is larger in W--R stars with smaller initial
mass $\mzams$ or with lower mass loss rate $\dot M$ due to higher
surface abundances of carbon and oxygen.
In the evolving W--R star the oscilation amplitude decreases with
decreasing stellar mass $M$ and for $M < 10M_\odot$ the sufficiently
small nonlinear effects allow us to calculate the integral of the
mechanical work $W$ done over the pulsation cycle
in each mass zone of the hydrodynamical model.
The only positive maximum on the radial dependence of $W$ is in the layers
with temperature of $T\sim 2\times 10^5\K$ where oscillations are excited
by the iron Z--bump $\kappa$--mechanism.
Radial oscillations of W--R stars with mass of $M > 10M_\odot$
are shown to be also excited by the $\kappa$--mechanism but
the instability driving zone is at the bottom of the envelope and
pulsation motions exist in the form of nonlinear running waves propagating
outward from the inner layers of the envelope.



\section*{Introduction}

During the core helium burning the structure of Population~I stars
with initial mass of $\mzams\ge 30M_\odot$ weakly depends on the
radial distribution of the mean molecular weight $\mu$.
This is due to the fact that radiation dominates in the internal energy,
so that the hydrostatic equilibrium is governed by the gravity and
by the gradient of radiation pressure whereas the gas pressure is
negligible.
Moreover, the opacity is mostly due to the Tomphson scattering
and radiative equilibrium above the convective core is almost
independent of the radial distribution of $\mu$.
Both these effects are strongest in stars with high effective temperature
($\Teff > 5\times 10^4\K$) and lead to the mass--luminosity relation
of W--R stars (Langer, 1989; M. Beech, R. Mitalas, 1992).

Together with the mass--luminosity relation
(Maeder, 1983; Doom et al., 1986; Maeder, 1987, Maeder, Meynet, 1987)
W--R stars obey the mass--radius relation
(Langer, 1989; Schaerer, Maeder, 1992)
and therefore the sound travel time between the stellar center and the
stellar surface can be considered approximately as a function of the
stellar mass $M$.
In recent years W--R stars were found to be unstable against radial
oscillations
(Glatzel et al., 1999; Fadeyev, Novikova, 2003; 2004; Dorfi et al., 2006;
Fadeyev, 2007; Fokin, Tutukov, 2007; Fadeyev, 2008),
so that one might assume that evolution of W--R stars is accompanined
by decrease of the period of their radial oscillations $\Pi$.
However one should bear in mind that if the pulsational instability
is due to the iron Z--bump $\kappa$--mechanism
(Dorfi et al., 2006; Fadeyev, 2008) then both the amplitude growth rate
and the amplitude of radial oscillations depend on the chemical composition
of the outer layers of the star.
Such a dependence is still unclear and
in the present paper we consider the pulsational properties of W--R stars
with various surface abundances of helium, carbon and oxygen.

The chemical composition of outer layers of the evolving star depends on
the both initial mass $\mzams$ and mass loss rate $\dot M$.
In this paper we consider the Population~I stars with initial masses
ranging within $70M_\odot\le\mzams\le 130M_\odot$.
To evaluate the effect of uncertainties in empirical estimates of mass
loss rates the evolutionary computations were carried out under various
assumptions on $\dot M$.
Some stellar models corresponding to the W--R phase were used as initial
conditions in hydrodynamic computations of nonlinear radial stellar
oscillations.
In the present paper we continue our previous studies (Fadeyev, 2007; 2008)
but now we use more recent data on the mass loss rate during helium burning
that are most important for the structure of W--R stars.

\section*{Method of computations}

Calculations of stellar evolution and stellar pulsations were done with
methods that in general were described in the previous paper (Fadeyev, 2007),
so that below only a few important changes are noted.

The mass loss rate during hydrogen burning and at the initial phase
of helium burning $\dmh$ was calculated according to
Nieuwenhuijzen and de Jager (1990).
The mass loss rate at the later stages of evolution $\dmhe$ was calculated
using the empirical formula by Nugis and Lamers (2000)
which takes into account both
clumping of the stellar wind and dependence of the mass loss rate on
surface abundances of helium and heavier elements.
In computations of stellar evolutuion the transition from $\dmh$ to $\dmhe$
was done during the initial stage of helium burning when
the effective temperature of the contracting star reaches $\Teff=10^4\K$.
At this point of the evolutionary track the difference between
$\dmh$ and $\dmhe$ is about a few percent.
For example, for $\mzams=80M_\odot$ and $\dot M = \dmh$
the star reaches the effective temperature $\Teff = 10^4\K$
when its mass is about $M=37M_\odot$ and surface abundances of
hydrogen and helium are $X(\hydr)\approx 0.1$ and $X(\hel)\approx 0.88$,
respectively.
The mass loss rates evaluated with formulae by
Nieuwenhuijzen and de Jager (1990) and by Nugis and Lamers (2000) are
$\dmh=9.3\times 10^{-5} M_\odot/\mathrm{yr}$ and
$\dmhe=9.6\times 10^{-5} M_\odot/\mathrm{yr}$.
It should be noted that during such a short phase
of contraction of the star the evolutionary track crosses
the H--R diagram so fast that
variation of the threshold effective temperature by a factor of
two around $\Teff=10^4\K$ does not affect perceptibly the
later evolution of the stars.

Thus, in calculations of stellar evolution the mass loss rate was
determined as
\begin{equation}
\dot M =
\left\{
\begin{array}{ll}
\fh \dmh   , & (\mbox{hydrogen burning and initial helium burning}) ,
\\[4pt]
\fhe \dmhe , & (\mbox{helium burning at\ } \Teff\ge 10^4\K) ,
\end{array}
\right.
\end{equation}
where constant factors $1 < \fh < 2$ � $0.5 < \fhe < 3$ were introduced
in order to evaluate the dependence of the results of evolution
calculations on uncertainties in mass loss rates.

In previous papers (Fadeyev, 2007; 2008) the thermodynamic functions
of the gas with temperature of $T < 10^7\K$ were calculated using the
OPAL equation of state data (Rogers et al., 1996).
Unfortunately, these data are not quite appropriate at late
stages of evolution when abundances of elements heavier than helium
substantially increase in the outer layers of the star.
In the present study we computed the tables of thermodynamic
quantities for about three dozen compositions, so that
both stellar evolution and pulsational instability can be
selfconsistently calculated up to the core helium
exhaustion when the stellar matter consists mostly of carbon
and oxygen.
Computing the tables of thermodynamic quantities we assumed that
heavy elements are carbon, nytrogen, oxygen and neon.
ZAMS abundances of these elements were taken from Rodgers et al. (1996).

The thermonuclear reaction network was extended to several dozen
isotopes from hydrogen ${}^1\mathrm{H}$ to nickel ${}^{56}\mathrm{Ni}$.
However in comparison with previous computations the larger number
of reactions did not affect significantly the results of evolution
calculations.
This is due to the fact that evolutionary calculations were
terminated just after the helium exhaustion when
the energy generation is due mostly to reactions
of the tripple--alpha process and
$\alpha({}^{12}\mathrm{C},\gamma){}^{16}\mathrm{O}$.

\section*{Mass--luminosity relation}

Domination of radiation in the helium burning phase
leads to convergence of the evolutionary tracks of massive stars
on the H--R diagram.
In the upper panel of Fig.~1 are shown the parts of two evolutionary
tracks with initial masses $\mzams=80M_\odot$ and $\mzams=130M_\odot$,
both of them being computed with $\fh=\fhe=1$.
At luminosity $\log L/L_\odot = 5.9$
masses and effective temperatures of these stars
($M\approx 25M_\odot$, $\Teff\approx 8.2\times 10^4\K$)
differ from one another by $2\%$ and $0.1\%$, respectively.
\begin{figure}
\centerline{\includegraphics[width=9cm]{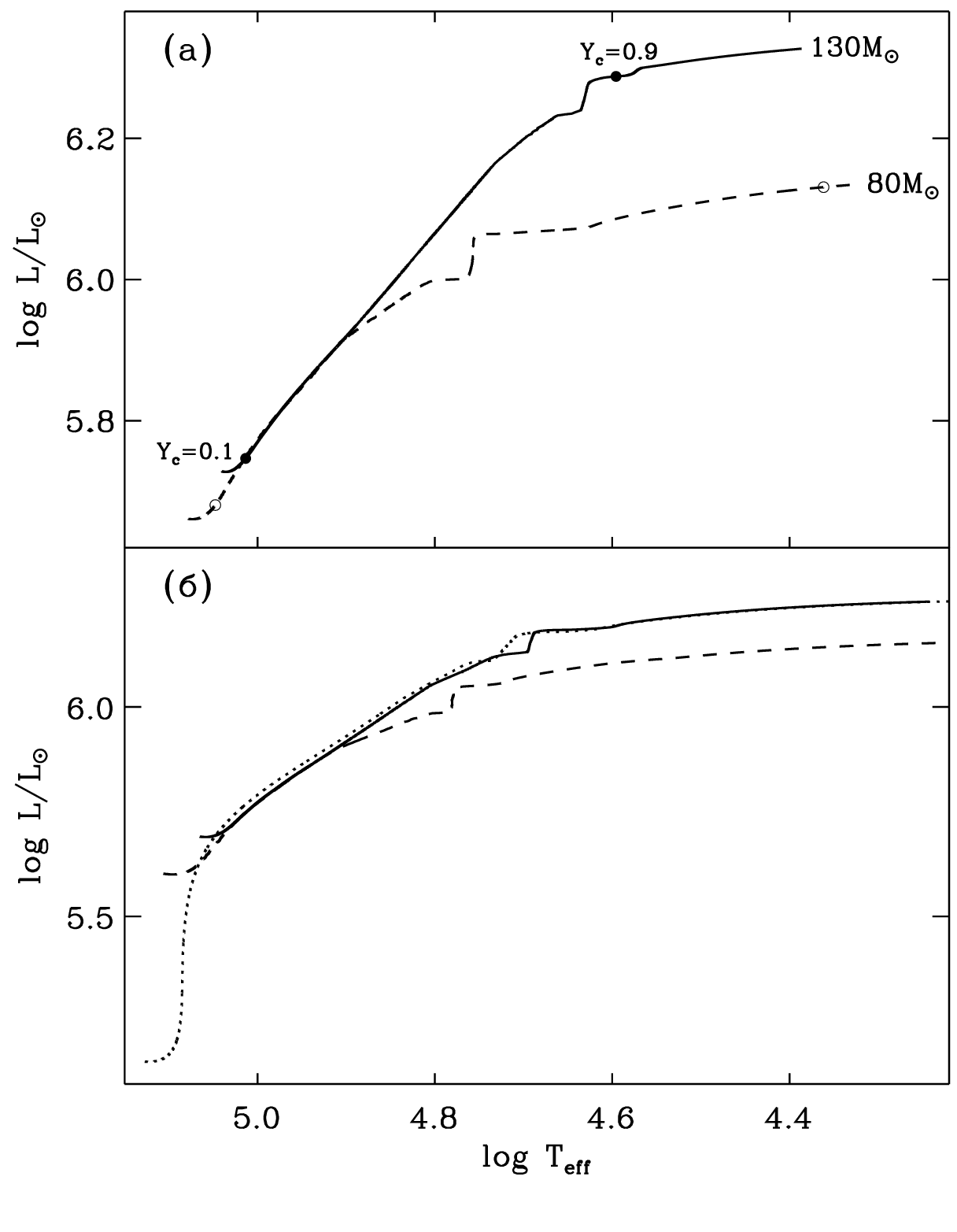}}
\caption{Upper panel: the parts of evolutionary tracks on the
H--R diagram for initial masses
$\mzams=130M_\odot$ (solid line) and $\mzams=80M_\odot$
(dashed line) with $\fh=\fhe=1$.
The circles indicate the models with central helium abundance
$Y_c=0.9$ and $Y_c=0.1$.
Lower panel: The parts of evolutionary tracks of the star
$\mzams=100M_\odot$ with $\fh=\fhe=1$ (solid line),
$\fh=2$, $\fhe=1$ (dashed line) and $\fh=1$, $\fhe=2$
(dotted line).
}
\label{fig1}
\end{figure}

W--R stars with close values of the stellar mass $M$, luminosity $L$
and radius $R$
but with different initial masses $\mzams$ have significantly
different radial distributions of the mean molecular weight.
This is illustrated in the upper panel of Fig.~2 where for two models
of W--R stars with mass of $M=25M_\odot$
($\mzams=80M_\odot$ and $\mzams=130M_\odot$) the
radial dependencies of the helium abundance $X(\hel)$ are plotted as
a functions of the Lagrangean mass coordinate $M_r$.
The lower helium abundance and the higher carbon abundance
in outer layers of the star with larger initial mass $\mzams$
is due to the more efficient helium burning during the
preceding stellar evolution.
Homogeneity of the chemical composition within the large mass
fraction of the star is due to the large extention of
convective cores.

\begin{figure}
\centerline{\includegraphics[width=7.2cm]{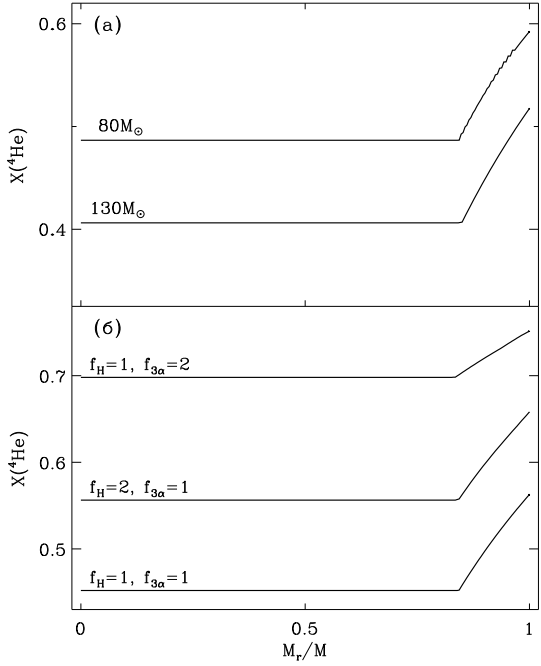}}
\caption{
Relative mass abundance of helium $X(\hel)$ as a function
of the Lagrangean mass coordinate $M_r$ in W--R stars with mass
of $M=25M_\odot$.
Upper panel: the case of mass loss $\fh=\fhe=1$.
Numbers at the curves indicate the initial stellar mass $\mzams$.
Lower panel: three different cases of mass loss for $\mzams=100M_\odot$.
}
\label{fig2}
\end{figure}

Dependence of the evolution of W--R stars on the mass loss rate
is illustrated in the lower panel of Fig.~1 where three evolutionary tracks
of the star with initial mass of $\mzams=100M_\odot$ are plotted for
three different cases of the mass loss:
($\fh=\fhe=1$), ($\fh=2$, $\fhe=1$) and ($\fh=1$, $\fhe=2$).
Doubling the mass loss rate $\dmh$ during hydrogen burning
leads to decrease of the luminosity of the W--R star with mass
of $M=25M_\odot$ by about $\approx 3\%$, whereas the core helium
abundance increases by $\Delta Y_c\approx 0.1$.

The most important role in the structure of W--R stars
belongs to the mass loss rate during the helium burning.
For example, doubling $\dmhe$ leads to decrease of the luminosity
in the W--R star with mass of $M=25M_\odot$ by $\approx 9\%$ and
at bthe same time to increase of the core helium abundance by
$\Delta Y_c\approx 0.25$.
Dependence of the radial distribution of helium abundance $X(\hel)$
on the mass loss rate $\dmhe$ is shown in the lower panel of Fig.~2.
Moreover, increase of $\dmhe$ is accompanied by the slower growth
of the central temperature, so that the end of core helium exhaustion
occurs at the lower stellar mass and on the H--R diagram the evolutionary
track extends to lower luminosities.

Convergence of the evolutionary tracks on the H--R diagram implies
the existence of the correlation between the stellar mass $M$
and the stellar luminosity $L$.
Fig.~3 shows the mass--luminosity diagram with two pairs of evolutionary
tracks for two initial stellar masses ($\mzams=90M_\odot$, $\mzams=130M_\odot$)
and two cases of mass loss:
($\fh=1$, $\fhe=1; \fh=2$, $\fhe=1$).
At the final point of each track plotted in Fig.~3 the central helium abundance
is $Y_c\approx 10^{-3}$.
It is clearly seen that during the significant part of the helium burning
phase the stellar mass and luminosity are nearly related by the power
dependence.
The position of the initial point of the power dependence changes
with initial stellar mass and mass loss rate.
For example, the central helium abundance at the initial point
ranges from $Y_c=0.85$ for $\mzams=70M_\odot$ to
$Y_c=0.94$ for $\mzams=130M_\odot$.
For smaller $\mzams$ or larger $\dmh$ the helium burning occurs
at the lower central temperature, so that on the mass--luminosity
diagram the part of the track with power dependence moves to
smaller values of $M$ and $L$.

\begin{figure}
\centerline{\includegraphics[width=10cm]{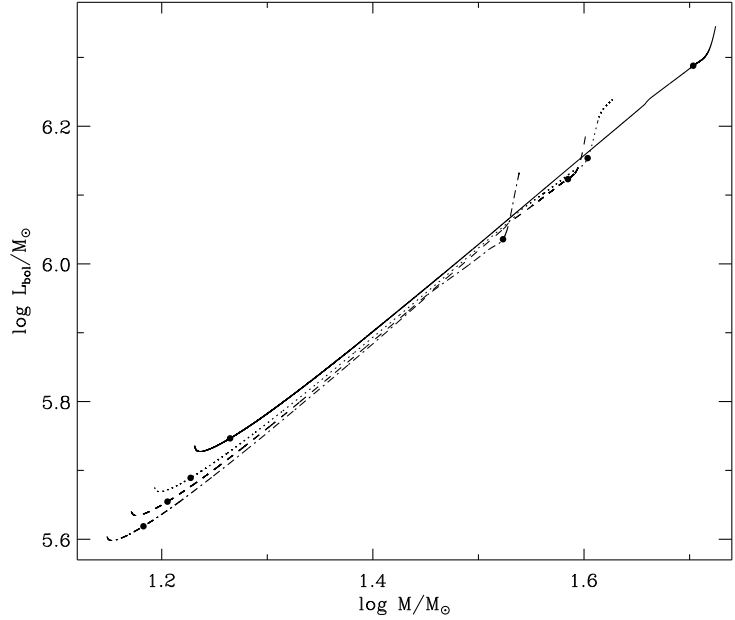}}
\caption{Mass--luminosity relation for $\fh=1$ and $\fh=2$
at initial masses
$\mzams=130M_\odot$ (solid and dashed lines) and
$\mzams=90M_\odot$ (dotted and dashed--dotted lines).
For all models $\fhe=1$.
}
\label{fig3}
\end{figure}

The parts of evolutionary tracks that can be approximated by the power
dependence on the mass--luminosity diagram are displayed in Fig.~4.
The maximum deviation of each track from the
power dependence is
$\dL=\max|\Delta\log L| \approx 0.01$ for $0.5\le\fhe\le 2$ and
$\dL \approx 0.03$ for $\fhe=3$.
The linear fit on the $(\log M,\log L)$ plane for models with
$70M_\odot\le\mzams\le 130M_\odot$, $1\le\fh\le 2$ and $0.5\le\fhe\le 3$
shown in Fig.~4 is given by
\begin{equation}
\label{wrml1}
\log L/L_\odot = 3.675 + 1.568 \log M/M_\odot .
\end{equation}

\begin{figure}
\centerline{\includegraphics[width=10cm]{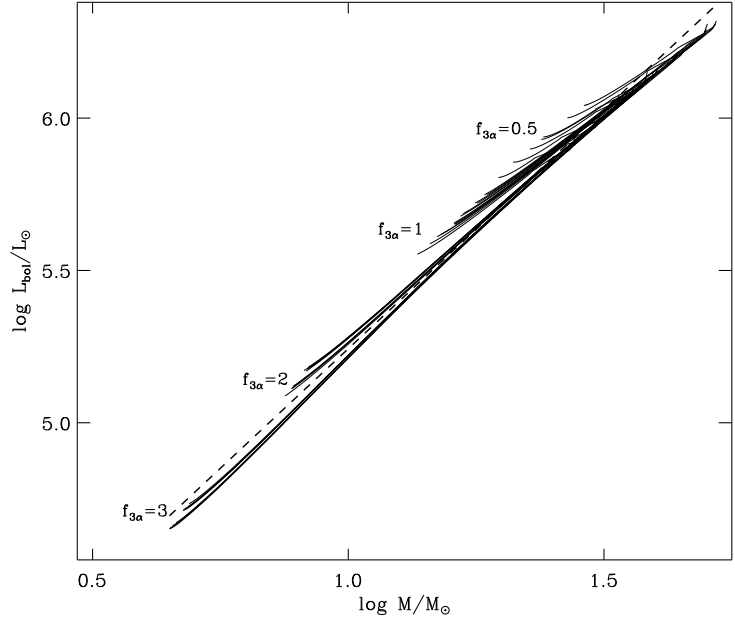}}
\caption{Mass--luminosity relation of W--R stars with
$70M_\odot\le\mzams\le 130M_\odot$, $1\le\fh\le 2$ and
$0.5\le\fhe\le 3$
The dashed line shows relation (ref{wrml1}).
}
\label{fig4}
\end{figure}

The maximum deviation of the parts of evolutionary tracks from
relation (\ref{wrml1}) is $\dL\le 0.11$ and its large value
is due to the dependence of the coefficients of this relation
on the mass loss rate $\dmhe$.
Therefore,  more exact approximation can be obtained for fixed
values of $\fhe$:
\begin{equation}
\label{wrml2}
\log L/L_\odot = \left\{
\begin{array}{lll}
 4.236 + 1.212 \log M/M_\odot , \quad & (\fhe = 0.5, & \dL\le 0.06) ,
\\[4pt]
 4.099 + 1.282 \log M/M_\odot , & (\fhe =   1, & \dL\le 0.03) ,
\\[4pt]
 3.821 + 1.454 \log M/M_\odot , & (\fhe =   2, & \dL\le 0.02) ,
\\[4pt]
 3.632 + 1.580 \log M/M_\odot , & (\fhe =   3, & \dL\le 0.04) .
\end{array}
\right.
\end{equation}
It should be noted that the mass interval within which
formulae (\ref{wrml2}) can be applied
depends on the initial mass $\mzams$ as well as on parameters $\fh$ and $\fhe$.
The stellar mass $M$ and the central helium abundance $Y_c$
corresponding to limits of these intervals are given in Table~1.
The last column of Table~1 gives the evolution time
$t_\mathrm{ev}$ for this interval.

\begin{table}
\caption{Limiting values of stellar mass $M$ and central helium abundance $Y_c$
in the mass--luminosity relations (\ref{wrml2}).
}
\begin{center}
\begin{tabular}{r|r|r|r|r|r|r|r}
\hline
$\fhe$ & $\fh$ & $\mzams/M_\odot$\quad & \multicolumn{2}{|c|}{$M/M_\odot$} & \multicolumn{2}{|c|}{$Y_c$} & $t_\mathrm{ev}$, $10^6$~yr�\\
\hline
 0.5 &  1  &   80  &  36.2  &  24.1  &  0.85  &  0.02  &  0.29  \\
%      &     &  100  &  44.1  &  26.8  &  0.90  &  0.03  &  0.28  \\
     &     &  120  &  50.1  &  28.9  &  0.93  &  0.05  &  0.27  \\
%      &  2  &   70  &  27.4  &  19.7  &  0.78  &  0.04  &  0.28  \\
     &  2  &   80  &  31.1  &  21.0  &  0.86  &  0.02  &  0.31  \\
%      &     &  100  &  35.5  &  22.6  &  0.90  &  0.03  &  0.30  \\
     &     &  120  &  38.4  &  23.8  &  0.92  &  0.03  &  0.30  \\
 1.0 &  1  &   80  &  36.1  &  16.6  &  0.88  &  0.09  &  0.28  \\
%      &     &  100  &  44.0  &  17.9  &  0.92  &  0.13  &  0.26  \\
     &     &  120  &  50.1  &  18.7  &  0.95  &  0.14  &  0.26  \\
%      &  2  &   70  &  27.8  &  13.6  &  0.89  &  0.06  &  0.32  \\
     &  2  &   80  &  31.0  &  14.4  &  0.90  &  0.08  &  0.30  \\
%      &     &  100  &  35.4  &  15.2  &  0.93  &  0.09  &  0.30  \\
     &     &  120  &  38.2  &  16.1  &  0.94  &  0.12  &  0.28  \\
 2.0 &  1  &   80  &  35.6  &   8.5  &  0.88  &  0.08  &  0.36  \\
%      &     &  100  &  43.9  &   8.3  &  0.93  &  0.06  &  0.38  \\
     &     &  120  &  49.9  &   8.2  &  0.95  &  0.05  &  0.41  \\
     &  2  &   80  &  31.0  &   7.7  &  0.92  &  0.08  &  0.38  \\
     &     &  120  &  38.3  &   7.8  &  0.95  &  0.07  &  0.40  \\
 3.0 &  1  &   80  &  36.2  &   4.9  &  0.90  &  0.01  &  0.54  \\
%      &     &  100  &  43.6  &   4.8  &  0.93  &  0.01  &  0.56  \\
     &     &  120  &  50.4  &   4.7  &  0.96  &  0.01  &  0.58  \\
     &  2  &   80  &  31.2  &   4.5  &  0.94  &  0.01  &  0.59  \\
     &     &  120  &  38.5  &   4.6  &  0.96  &  0.02  &  0.58  \\
\hline
\end{tabular}
\end{center}
\end{table}

When one considers the stellar mass $M$ as a parameter
one should bear in mind that the stellar evolution becomes slower
as $M$ decreases.
In Fig.~5 for $\mzams=100M_\odot$ and $\fh=1$ are shown
the plots of the life time $\Delta t_\mathrm{ev}$
within the mass interval $[M-\Delta M, M]$, where $\Delta M=0.5M_\odot$.
The plots shown in Fig.~5 weakly depend on $\mzams$ and $\fh$.
The growth of the life time $\Delta t_\mathrm{ev}$ with increasing $\fhe$
is due to the slower increase of the central temperature and
slower conversion of helium into carbon.
For initial stellar mass $\mzams=100M_\odot$ and mass loss rate
during hydrogen burning $\fh=1$ the second half of the life time
correspond to W--R masses
$M<33M_\odot$, $M<25M_\odot$, $M<12M_\odot$ and $M<7M_\odot$
for $\fhe=0.5$, 1, 2 and 3, respectively.

\begin{figure}
\centerline{\includegraphics[width=8.5cm]{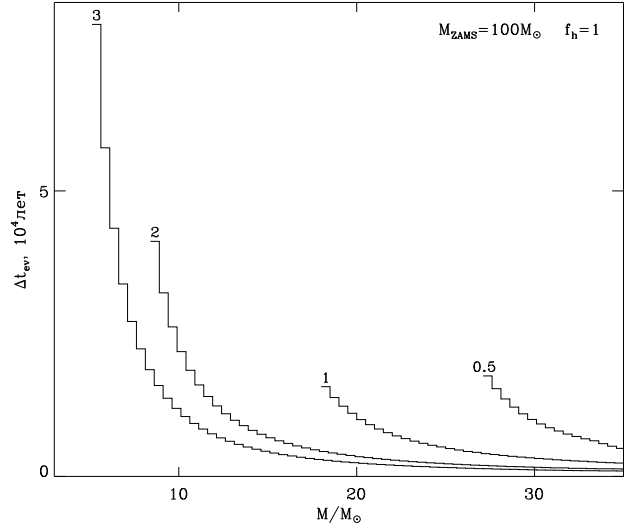}}
\caption{The life time $\Delta t_\mathrm{ev}$ of W--R stars
within the mass interval $\Delta M=0.5M_\odot$
as a function of the stellar mass $M$.
The initial mass is $\mzams=100M_\odot$ and $\fh=1$.
The numbers at the curves indicate correspondig values of $\fhe$.
}
\label{fig5}
\end{figure}

\section*{Nonlinear radial oscillations}

During evolution of the W--R star the relative radius of the stellar core increases,
so that both the sound travel time from the center to the surface
and the period of radial oscillations gradually decrease.
At the same time the instability excitation zone moves closer to
the stellar surface and the amplitude of oscillations decreases
due to diminishing mass of the driving zone.
Decrease of the amplitude and the period of radial oscillations in
evolving W--R stars is shown in Fig.~6 where
for stars with initial masses $\mzams=90M_\odot$ and $\mzams=120M_\odot$
are given the plots of the maximum expansion velocity of the outer boundary
of the hydrodynamic model $\umax$ expressed in units of the local escape velocity
$\vesc$ and the period of radial oscillations $\Pi$.

\begin{figure}
\centerline{\includegraphics[width=7.4cm]{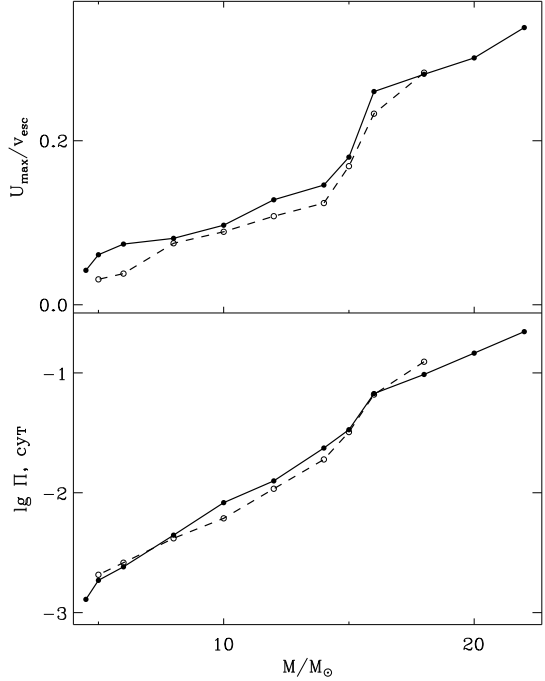}}
\caption{
The ratio of the maximum expansion velocity of the
outer boundary $\umax$ to the local escape velocity $\vesc$
(upper panel) and the period of radial oscillations $\Pi$ in days
(lower panel)
as a function of the stellar mass $M$.
In solid and dashed lines are shown the evoltuionary
sequences with
$\mzams=120M_\odot$ and $\mzams=90M_\odot$, respectively.
}
\label{fig6}
\end{figure}

The growth time of radial oscillations is of the order of the
stellar dynamic time scale, so that oscillations of W--R stars
are stronly nonadiabatic.
In contrast to many other radially pulsating stars
oscillations of W--R stars cannot be described in terms
of the standing wave since the kinetic energy of the pulsating
envelope only once per period passes the minimum and the maximum.
Radial oscillations of W--R stars should be considered as
nonlinear running waves propagating from the envelope bottom
to the outer boundary.
That is why in W--R stars with mass of $M > 15M_\odot$
the pulsation constant ($Q\ge 0.1$~day) is
significantly larger in comparison with pulsation constants
of stars radially pulsating in the form of standing waves.
The only exception is W--R stars with mass of $M < 10M_\odot$
where the small--amplitude radial oscillations can be approximately
represented by nonadiabatic standing waves.

In the stellar mass range of $4.5M_\odot\le M\le 20M_\odot$
the pulsation constant of W--R stars can be approximately
written as
\begin{equation}
\log Q = -2.6 + 0.1 M_\odot .
\end{equation}

Properties of some hydrodynamic models are listed in Table~2.
The mass of outer pulsating layers is negligible in comparison
with the total mass of the star, so that the
abundances of helium $X(\hel)$, carbon $X(\carb)$ and oxygen
$X(\ox)$ are constant through the envelope.
The mean pulsation period $\Pi$ was evaluated using the
discrete Fourier transform of the kinetic energy of
the oscillating envelope within the time interval
$10^2\lesssim t/\Pi\lesssim 10^3$.
However, strictly speaking, the definition of the period of radial oscillations
can be applied to W--R stars with rather low stellar masses because
for $M > 15M_\odot$ the frequency of oscillations at the bottom
of the envelope becomes somewhat higher than that of the outer layers.
In Fig.~7 are shown the power spectra of the velocity of gas
in the outer ($r\approx R$) and the inner ($r\approx 0.79R$)
layers of the envelope of the W--R star with mass of $M=16M_\odot$.
Contribution of short--period oscillations in the inner layers
becomes significant in stars with $M > 20M_\odot$
because they affect the radiative flux emerging from the
outer boundary and the period of light variatins
becomes shorter than that of hydrodynamic motions in the
outer layers of the pulsating envelope.

\begin{figure}
\centerline{\includegraphics[width=9cm]{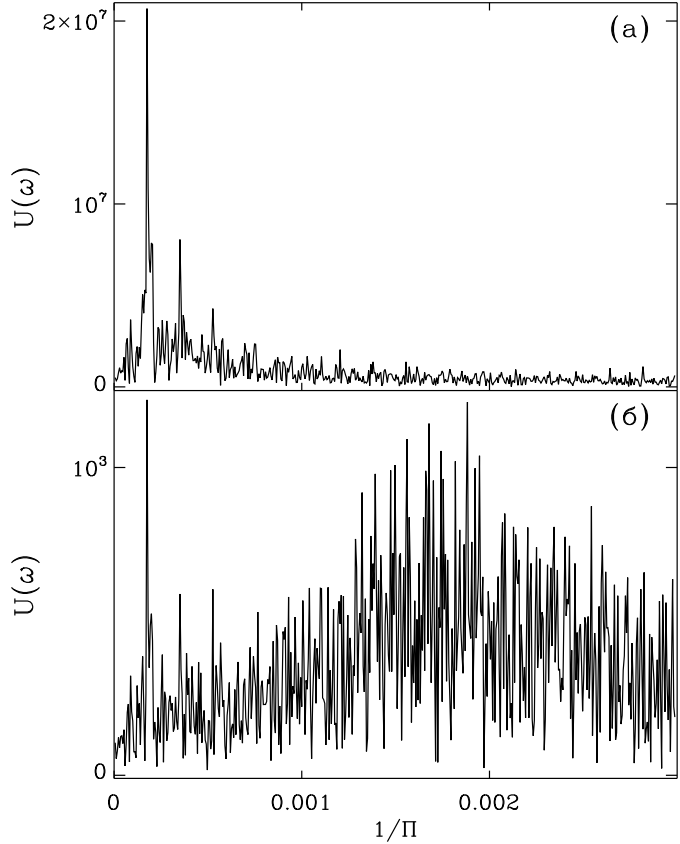}}
\caption{The power spectrum of velocity $U(\omega)$ in the outer
(upper panel) and the inner (lower panel) layers of the W--R star
with mass $M=16M_\odot$.
}
\label{fig7}
\end{figure}

\begin{table}
\caption{Properties of some hydrodynamic models of W--R stars.}
\begin{tabular}{r|r|r|r|r|r|r|l|l|l}
\hline
$\mzams/M_\odot$ & $\fhe$ & $M/M_\odot$ & $\log L/L_\odot$ & $X(\hel)$ & $X(\carb)$  & $X(\ox)$ & $\Pi$~day & $\umax/\vesc$ & $\bar{R}/R$\\
\hline
 90 & 2 &  15   & 5.541 & 0.639  &  0.311  &  0.031 & 0.0525 & 0.21 & 2.1  \\
    &   &  10   & 5.272 & 0.467  &  0.436  &  0.078 & 0.0085 & 0.13 & 1.1 \\
    & 3 &  18   & 5.660 & 0.786  &  0.186  &  0.009 & 0.124  & 0.30 & 3.4 \\
    &   &  15   & 5.524 & 0.745  &  0.223  &  0.014 & 0.0321 & 0.17 & 1.5 \\
    &   &  10   & 5.221 & 0.635  &  0.316  &  0.030 & 0.0061 & 0.09 & 1.1  \\
    &   &   5   & 4.733 & 0.368  &  0.502  &  0.111 & 0.0021 & 0.03 & 1.0  \\
120 & 2 &  15   & 5.553 & 0.634  &  0.315  &  0.033 & 0.0652 & 0.25 & 2.5 \\
    &   &  10   & 5.289 & 0.461  &  0.439  &  0.081 & 0.0107 & 0.14 & 1.2  \\
    & 3 &  22   & 5.772 & 0.827  &  0.148  &  0.006 & 0.221  & 0.34 & 6.4  \\
    &   &  20   & 5.712 & 0.807  &  0.166  &  0.008 & 0.146  & 0.30 & 3.8  \\
    &   &  15   & 5.528 & 0.742  &  0.225  &  0.014 & 0.0336 & 0.18 & 1.5  \\
    &   &  10   & 5.248 & 0.632  &  0.319  &  0.031 & 0.0083 & 0.11 & 1.1  \\
    &   &   5   & 4.743 & 0.369  &  0.501  &  0.111 & 0.0018 & 0.06 & 1.0  \\
\hline
\end{tabular}
\end{table}

As is seen in Table~2 the pulsational properties of W--R stars
depend mostly on the mass loss rate $\dmhe$ whereas the role
of the both initial stellar mass $\mzams$ and mass loss rate
during hydrogen burning $\dmh$ is significantly weaker.
This is due to the fact that variations of $\dmhe$ are
accompanied by significant changes of the both stellar luminosity
and surface abundances of carbon and oxygen.
In particular, increase of the mass loss rate during helum burning
leads to the smaller stellar luminosity and therefore to the
smaller nonadiabaticity of stellar pulsations.

An important consequence of nonlinear radial stellar oscillations
is the increase of the mean radius of pulsating layers of the gas.
This effect is illustrated by Table~2 where the
last column gives the mean radius of the
outer boundary of the hydrodynamic model $\bar R$ expressed in
units of the initial equilibrium radius $R$.

\section*{Mechanism of pulsational instability}

The elementary spherical layer of gas
contributing into excitation of pulsational instability
performs the positive mechanical work during the
pulsation cycle, that is the integral of mechanical work is
positive : $W = \displaystyle\oint PdV > 0$,
where $P$ is the total pressure and $V$ is the specific volume.
Unfortunately, for hydrodynamical models of W--R stars
exact calculation of the radial dependence of the mechanic work $W$
is impossible because of strongly nonlinear radial oscillations.
The only exception is W--R stars with mass of $M \le 10M_\odot$
where pulsation motions are characterized by a
good repetition of pulsation cycles.
Formation of W--R stars with so small masses implies rather high
mass loss rates during the core helium burning phase ($\fhe\ge 3$).
Omitting discussion on possibility of the existence of such
W--R stars we consider their pulsational properties
because the results obtained can be generalized to more massive
W--R stars undergoing substantially smaller mass loss.

The upper panel of Fig.~8 shows radial dependences of the inetgral
of mecahnical work $W$ for two models of W--R stars with mass of
$M=6M_\odot$ and $M=8M_\odot$.
It is clearly seen that excitation of oscillations ($W>0$) occurs in the
outer layers of the stars and the maximum of $W$ moves to the
surface as the stellar mass $M$ decreases.
To clarify the origin of the
pulsational instability one should compare the radial dependence
of $W$ with that
of the amplitude of radiative flux variations.
To this end we consider the spectral density of luminosity
\begin{equation}
L_r(\omega) = \int\limits_{-\infty}^\infty L_r e^{i\omega t} dt
\end{equation}
at the angular frequency $\omega = 2\pi/\Pi$.
The spectral density $L_r(\omega)$ was computed in all Lagrangeam
mass zones using the discrete Fourier transform
within time intervals $t/\Pi\lesssim 10^3$.
The lower panel of Fig.~8 shows two radial dependences of $L_r(\omega)$
that are normalized to the surface value.

\begin{figure}
\centerline{\includegraphics[width=8cm]{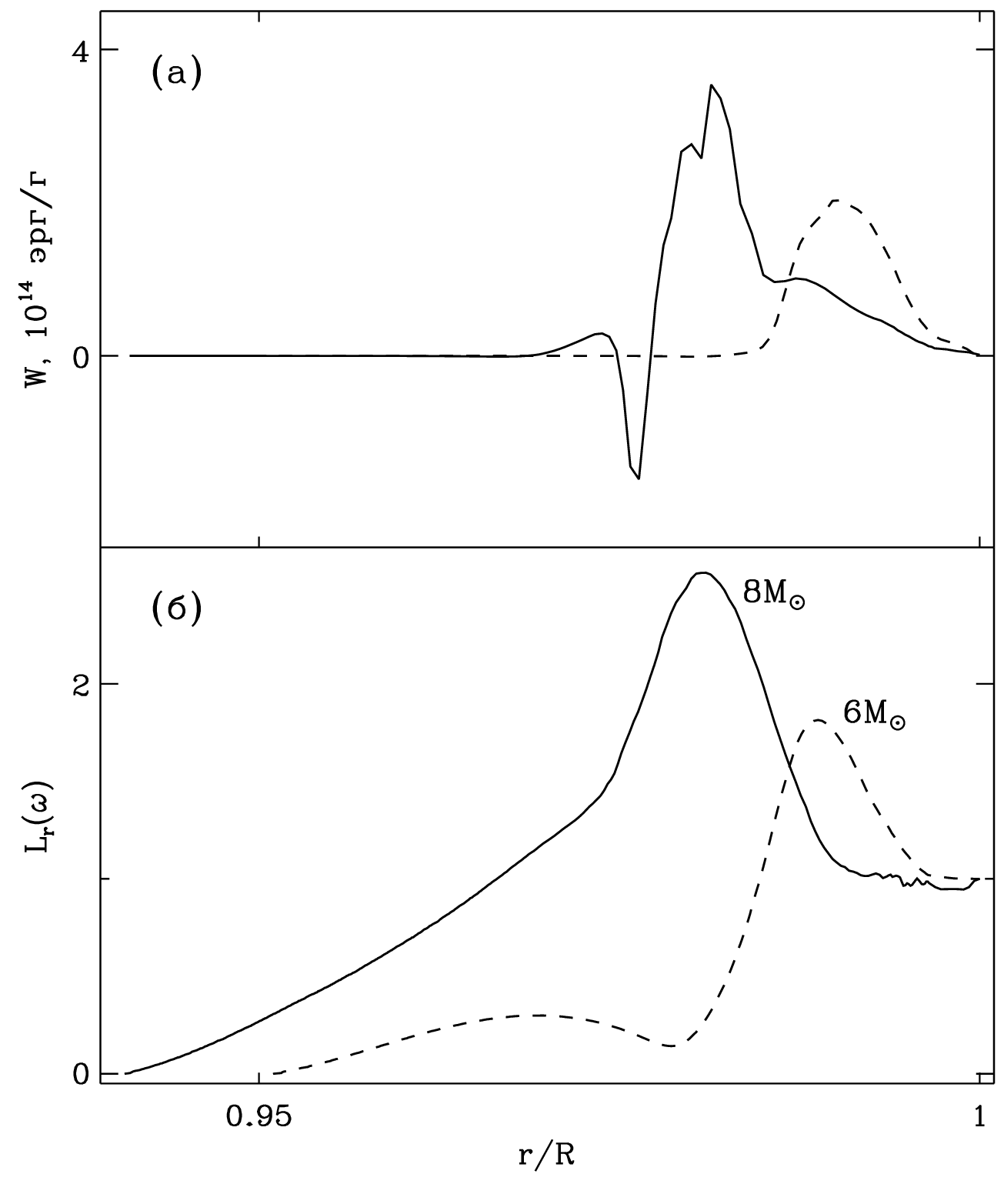}}
\caption{Radial dependences of the mechanical work over the
pulsation cycle $W$ (upper panel) and the normalized
spectral density of luminosity
$L_r(\omega)$ (lower panel) for models of W--R stars with masses
$M=6M_\odot$ and $M=8M_\odot$.
}
\label{fig8}
\end{figure}

Coincidence of radial coordinates of maxima of $W$ and
$L_r(\omega)$ (see Fig.~8) allows us to conclude
that excitation of oscillations is due to the interaction of
radiative flux with gas of the envelope.
In particular, for $W > 0$ it is necessary that
the gas absorbs radiation at maximum compression
and becomes more transparent at maximum expansion.
Fig.~9 shows the plots of variations of the gas density
$\rho$ and opacity $\kappa$ in the mass zone with maximum
mechanical work $W$ of the model of W--R stars with mass $M=8M_\odot$.
It is clearly seen that the positive mechanical work $W$
is due to the $\kappa$--mechanism.
The average temperature of gas in this zone is
$T\sim 2\times 10^5\K$ and the positive temperature
derivative of opacity
$(\partial\ln\kappa/\partial\ln T)_\rho > 0$
is due to the iron Z-bump.

\begin{figure}
\centerline{\includegraphics[width=9cm]{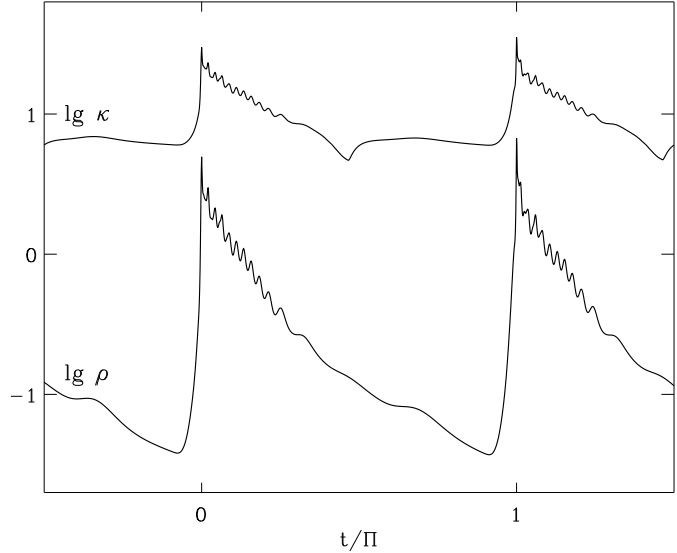}}
\caption{Variations of the gas density $\rho$ and opacity $\kappa$
in the vicinity of the maximum of mechanical work $W$
in the model of the W--R star with mass $M=8M_\odot$.
The plots of $\rho$ and $\kappa$ are arbitrarily shifted
along the vertical axis.
}
\label{fig9}
\end{figure}

Fig.~10 shows the plots of radial dependencies of the gas density $\rho$
and the spectral density of luminosity $L_r(\omega)$
for outer layers of W--R stars with masses $12M_\odot\le M\le 18M_\odot$.
The narrow maximum of $L_r(\omega)$ (see the lower panel of Fig.~10)
corresponds to layers with agerage temperature $T\sim 2\times 10^5\K$
where the gas density $\rho$ and opacity $\kappa$ reach the maximum
simultaneously, that is excitation of oscillations is also
due to the $\kappa$--mechanism.
In layers above the excitation zone
$\partial L_r(\omega)/\partial r \approx 0$,
that is there are neither excitation nor damping zones.
More massive W--R stars possess more extended envelopes with
smaller gradient of the gas density.
The excitation zone is at the bottom of the envelope and
radial oscillations exist as successive nonlinear waves
propagating from the outer boundary of the core to
the stellar surface.

\begin{figure}
\centerline{\includegraphics[width=8cm]{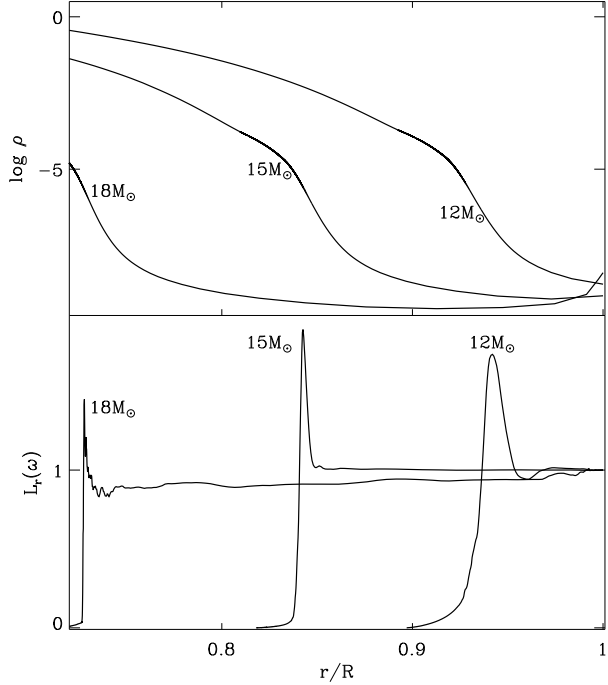}}
\caption{Radial dependencies of gas density $\rho$ (upper panel)
and normalized spectral density of luminosity $L_r(\omega)$ (lower panel)
in W--R stars with mass $M=12M_\odot$, $15M_\odot$ and $18M_\odot$.
}
\label{fig10}
\end{figure}

\section*{Conclusion}

As follows from results presented above
the mass and luminosity of Population~I massive stars
are related by the power dependence during the large part
of the helium burning phase.
Boundaries of the segment of the evolutionary track within which
the coefficients of the power dependence are constant depends on the
initial stellar mass $\mzams$ and the mass loss rate $\dot M$
during the preceding evolution.
At the initial point of the power dependence
the central helium abundance ranges from
$Y_c\approx 0.78$ ($\mzams=70M_\odot$, $\fhe=0.5$) to
$Y_c\approx 0.96$ ($\mzams=130M_\odot$, $\fhe=3$),
whereas the effective temperature is in the range from
$\Teff\approx 3\times 10^4\K$ ($\mzams=120M_\odot$) to
$\Teff\approx 5\times 10^4\K$ ($\mzams=70M_\odot$).
At the end of the power dependence the central helium
abundance ranges within $0.01\lesssim Y_c\lesssim 0.15$.
Thus, the limits of applicability of the mass--luminosity relation
depend on the both initial mass $\mzams$ and mass loss rate $\dot M$,
so that relations (\ref{wrml2}) should be used with
corresponding boundary values of the stellar mass $M$ listed
in Table~1.

In the present study we considered pulsational properties of
W--R stars as a function of three parameters: the initial stellar
mass $\mzams$, the mass loss rate during hydrogen burning $\dmh$
and the mass loss rate during the helium burning phase $\dmhe$.
However the most important is $\dmhe$ since its variations lead
to significant changes in surface abundances of helium, carbon
and oxygen.
In particular, at lower mass loss rate $\dmhe$
radial oscillations of W--R stars have larger amplitudes due to
higher surface abundances of carbon and oxygen.

As is seen from Fig.~6 and Table~2 the period of radial oscillations
is a sensitive indicator of the stellar mass and during evoulution
of the W--R star decreases from $\approx 5.5$~hr at $M=22M_\odot$ to
$\approx 2.6$~min at $M=5M_\odot$.
Thus, observational estimates of pulsation periods
could provide with a direct evaluation of the masses
of W--R stars.
For example, aAccording to Veen et al. (2002a, 2002b, 2002c)
the pulsation period of WR46 is $\Pi\approx 0.14$~day
and as follows from our hydrodynamical calculations
the stellar mass is $M\approx 20M_\odot$.

The present study is confined to hydrodynamical models of W--R
stars near their final stage of evolution and
such a choice are is to their longer life time.
However of great interest are more massive W--R stars
with higher luminosity and much stronger instability
similar to that of LBV stars.
The study of these objects will be presented in the
forthcoming paper.

\newpage
\section*{References}

\begin{enumerate}

\item M. Beech, R. Mitalas, Astron. Astrophys., \textbf{262}, 483 (1992).

\item P.M. Veen, A.M. van Genderen, K.A. van der Hucht, et al., Astron.Astrophys., \textbf{385}, 585 (2002�).

\item P.M. Veen, A.M. van Genderen, K.A. van der Hucht, et al., Astron.Astrophys., \textbf{385}, 600 (2002�).

\item P.M. Veen, A.M. van Genderen, K.A. van der Hucht, et al., Astron.Astrophys., \textbf{619}, 585 (2002�).

\item E.A. Dorfi, A. Gautschy, H. Saio, Astron.Astrophys., \textbf{453}, L35 (2006).

\item C. Doom, J.P. de Greve, C. de Loore, Astrophys. J., \textbf{303}, 136 (1986).

\item W. Glatzel, M. Kiriakidis, S. Chernigovskij, et al., MNRAS, \textbf{303}, 116 (1999).

\item N. Langer, Astron. Astrophys., \textbf{210}, 93 (1989).

\item A. Maeder, Astron. Astrophys., \textbf{120}, 113 (1983).

\item A. Maeder, Astron. Astrophys., \textbf{173}, 247 (1987).

\item A. Maeder, G. Meynet,  Astron. Astrophys., \textbf{182}, 243 (1987).

\item N. Nugis, H. J. G. L. M. Lamers, Astron. Astrophys., \textbf{360}, 227 (2000).

\item H. Nieuwenhuijzen and C. de Jager, Astron.Astrophys., \textbf{231}, 134 (1990).

\item F.J. Rogers, F.J. Swenson, and C.A. Iglesias, Astrophys.J., \textbf{456}, 902 (1996).

\item Yu.A. Fadeyev, M.F. Novikova, Ast.Let., \textbf{29}, 522 (2003).

\item Yu.A. Fadeyev, M.F. Novikova, Ast.Let., \textbf{30}, 707 (2004).

\item Yu.A. Fadeyev, Ast.Let., \textbf{33}, 692 (2007).

\item Yu.A. Fadeyev, Ast.Rep., in press (2008).

\item A.B. Fokin, A.V. Tutukov, Ast.Rep, \textbf{51}, 742 (2007).

\item D. Schaerer, A. Maeder, Astron. Astrophys., \textbf{263}, 192 (1992).

\end{enumerate}

\end{document}